\documentclass[review]{elsarticle}

\usepackage{lineno,hyperref}

\usepackage{subfigure}
\usepackage{multirow}
\usepackage{color,soul}
\usepackage{amsmath}
\usepackage{amsthm}
\usepackage{amssymb,bm}

\journal{Engineering Structures}

\begin{document}

\begin{frontmatter}

\title{Analytical and simplified models for dynamic analysis of short skew bridges under moving loads}

\author[UPM]{K. Nguyen\corref{mycorrespondingauthor}}
\ead{khanh@mecanica.upm.es}
\cortext[mycorrespondingauthor]{Corresponding author}

\address[UPM]{Group of Computational Mechanic, School of Civil Engineering, UPM, Spain }

\author[UPM]{J.M. Goicolea}
\ead{jose.goicolea@upm.es}

\begin{abstract}
Skew bridges are common in highways and railway lines when non
perpendicular crossings are encountered. The structural effect of skewness is an additional torsion on the bridge deck which may have a considerable effect, making its analysis and design more complex. In this paper, an analytical model following 3D beam theory is firstly derived in order to evaluate the dynamic response of skew bridges under moving loads. Following, a simplified 2D model is also considered which includes only vertical beam bending. The natural frequencies, eigenmodes and orthogonality relationships are determined from the boundary conditions. The dynamic response is determined in time domain by using the ``exact'' integration. Both models are validated through some numerical examples by comparing with the results obtained by 3D FE models. A parametric study is performed with the simplified model in order to identify parameters that significantly influence the vertical dynamic response of the skew bridge under traffic loads. The results show that the grade of skewness has an important influence on the vertical displacement, but hardly on the vertical acceleration of the bridge. The torsional stiffness really has effect on the vertical displacement when the skew angle is large. The span length reduces the skewness effect on the dynamic behavior of the skew bridge.
\end{abstract}


\end{frontmatter}


\section{Introduction}
Skew bridges are common in highways and railway lines when non perpendicular crossings are encountered. The structural effect of the skewness is an additional torsion on the bridge deck \cite{Kollbrunner1969, Manterola2006} which may have a considerable effect, making its analysis and design more complex. A large research effort using the analytical, numerical as well as experimental approaches have been made during the last decades in order to better understand the behavior of this type of bridge under the static and dynamic loadings. Special attention is given in researches related to the highway skew bridge subjected to earthquake loadings. In fact, the first work on this subject was reported in 1974 by Ghobarah and Tso \cite{Ghobarah1974}, in which a closed-form solution based on the beam model capable of capturing both flexural and torsional modes was proposed to study the dynamic response of the skewed highway bridges with intermediate supports. Maragakis and Jennings \cite{Maragakis1987} obtained the earthquake response of the skew bridge, modelling the bridge deck as a rigid body. Using the Finite Element (FE) Models, the so-called \textit{stick model} is firstly introduced by Wakefield et al. \cite{Wakefield1991}. The stick model consists of a beam element representing the bridge deck, rigid or flexible beam elements for the cap-beam and an array of translational and rotational springs for the substructure of the bridge. This type of model is then successfully used in the later works \cite{Meng2000,Meng2001,Nielson2007,Abdel-Mohti2008,Kaviani2012,Yang2015}. Despite its simplicity, the stick model can provide reasonably good approximations for the preliminary assessment. More sophisticated 3D models using the shell and beam elements are also proposed to study this subject \cite{Meng2000,Meng2002,Abdel-Mohti2008,Nouri2012,Deng2015,Mallick2015}. Regarding the behavior of the skew bridges under the traffic loads, the most of the work about this subject has been performed on the FE models using the combination of shell and beam elements and assisted by experimental testing \cite{Bishara1993,Helba1995,Khaloo2003,Menassa2007,Ashebo2007,He2012}. The FE models give a good approximation but require the end user more effort to introduce information in modelling the structure such as element types and sizes, dimension, material properties, connection types, etc. Therefore, its use is limited in determined case studies and challenged for a parametric study as Monte Carlo simulations or large number of case studies. A possible alternative is to develop an analytical solution that is able to capture the behavior of the skew bridge and to give a sufficient accuracy. The advantage of the analytical solution is that the data input is much simpler (general information of structure such as mass, span length, flexural and torsional stiffness) and, therefore, its use is more easy for the end user and ,of course, is able for parametric study. 

In this context, the main objective of this work is to derive an analytical solution based on the beam theory for the simply-supported skew bridge under the moving loads. After that, a simplified 2D model is proposed in order to assimilate the effect of the skewness of the support on the vertical vibration of the bridge. An "exact" integration in the time domain is used to solve the differential equations. Both models are validated through some numerical examples by comparing with the results obtained by 3D FE models. A parametric study is performed with the simplified model in order to identify parameters that significantly influence the vertical dynamic response of the skew bridge under traffic loads
\section{Formulation of problem}
\begin{figure}[h!]
	\centering
	\includegraphics[width=0.6\textwidth]{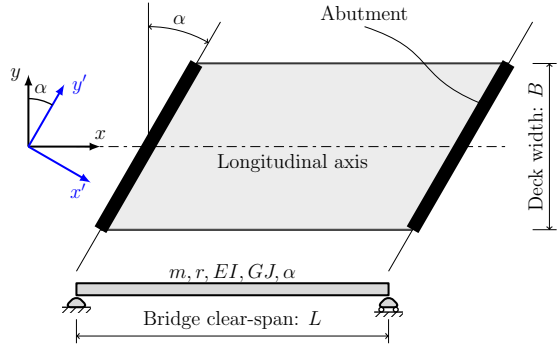}
	\caption{A simply-supported skew bridge: in plane view and bridge model's sketch}
	\label{fig.1}
\end{figure}

A simply-supported skew bridge as shown in Fig. \ref{fig.1} is considered to study in this work. The line of abutment support forms with the orthogonal line of the centreline an angle $\alpha$ defined as angle of skewness. The length of bridge is taken as the clear-span length $\mathrm{L}$. The bridge is idealized using following assumptions:
\begin{itemize}
	\item The bridge deck is modelled as 3D Euler-Bernoulli beam supported at the ends and has a linear elastic behavior.
	\item The bridge deck is very stiff in the horizontal XY plane, so the flexural deflection in Y direction will be neglected.
	\item The bending stiffness $\mathrm{EI}$, torsional stiffness $\mathrm{GJ}$ and mass per unit length $m$ are constant over the length $\mathrm{L}$.
	\item Warping and distortion effects in the torsion of the bridge deck is small enough to be neglected.
\end{itemize}

With these assumptions, the bending of the bridge in XZ plane and its twisting about the X  axis are the principal types of deformation of the bridge deck. The governing equations of motion for transverse and torsional vibration under transverse and torsional loads are:

\begin{subequations}
\begin{align}
	m\ddot{u} + c \dot{u} + EI \frac{\partial^4 u}{\partial x^4} = p(x,t) \label{eq.mov1} \\
	mr^2\ddot{\theta} - GJ \frac{\partial^2 \theta}{\partial x^2} = m_t(x,t) \label{eq.mov2}
\end{align}
\end{subequations}
where $r$ is the radius of gyration; $u(x,t)$ and $\theta(x,t)$ are the transverse deflection and torsional rotation of the bridge deck; $p(x,t)$ and $m_t(x,t)$ are the transverse and torsional loads applied on the bridge at distance $x$ and at time $t$, respectively. The external damping mechanism is introduced by the familiar term $c\dot{u}$ and is assumed to be proportional to the mass ($c=2m\xi_n\omega_n$)
\subsection{Natural frequencies and mode shapes}
\label{sec.2.1}
Using the modal superposition technique, the solution for free vibrations of the bridge deck can be decoupled into an infinite set of modal generalized coordinates and mode shapes as:

\begin{subequations}
	\begin{align}
		u(x,t) = \sum_{n=1}^{\infty} q_{n}(t) \phi_{n}(x) \label{eq.descomp1}\\
		\theta(x,t) = \sum_{n=1}^{\infty} p_{n}(t) \varphi_{n}(x) \label{eq.descomp2}
	\end{align}
	\label{eq.descomp}
\end{subequations}
in which $\phi_n(x)$ and $\varphi_n(x)$ the $n^{th}$ flexural and torsional mode shape, and $q_n(t)$ and $p_n(t)$ are the generalized flexural and torsional coordinates at $n^{th}$ mode shape and are assumed to be $e^{i\omega_nt}$. The governing equations for free vibrations can be rewritten for each mode of vibration as:   

\begin{subequations}
	\begin{align}
		\frac{1}{\phi_n(x)} \frac{\mathrm{d}^4 \phi_n(x)}{\mathrm{d} x^4} &= \frac{m\omega_n^2}{EI} \label{eq.diff1}\\
		\frac{1}{\varphi_n(x)} \frac{\mathrm{d}^2\varphi_n(x)}{\mathrm{d} x^2} &=- \frac{m r^2 \omega_n^2}{GJ}  \label{eq.diff2}
	\end{align}
	\label{eq.diff}
\end{subequations}
The solutions of the above equations can found in many textbooks on dynamic, and can be expressed in the following form:

\begin{subequations}
	\begin{align}
		\phi_n(x) & = C_{n,1} \sin(\beta_n x) + C_{n,2} \cos(\beta_n x) + C_{n,3} \sinh(\beta_n x) + C_{n,4} \cosh(\beta_n x) \label{eq.mode_shape1}\\
		\varphi_n(x) &= C_{n,5} \sin(\lambda_n x) + C_{n,6} \cos(\lambda_n x) \label{eq.mode_shape2}
	\end{align}
	\label{eq.mode_shpe}
\end{subequations}
where $\beta_n^4 = m\omega_n^2/EI$ and $\lambda_n^2 = mr^2\omega_n^2/GJ$; $C_{n,1},~C_{n,2},~C_{n,3},~C_{n,4},~C_{n,5},~C_{n,6}$ are six constants that are determined by the boundary conditions.

The boundary conditions for the problem are shown in Fig. \ref{fig.1}. The bridge is simply-supported at the ends by abutments. Therefore, at the support lines, there are not the vertical displacement ($u(0,t)=u(L,t)=0$), rotation about the $X'$ axis ($\theta_{x'}(0,t)=\theta_{x'}(L,t)=0$) and bending moment in $X'$ axis ($M_{y'}(0,t)=M_{y'}(L,t)=0$). Using the change of coordinates as shown in Fig. \ref{fig.2}, the following relationships are obtained:

\begin{figure}
	\centering
	\includegraphics[width=0.4\textwidth]{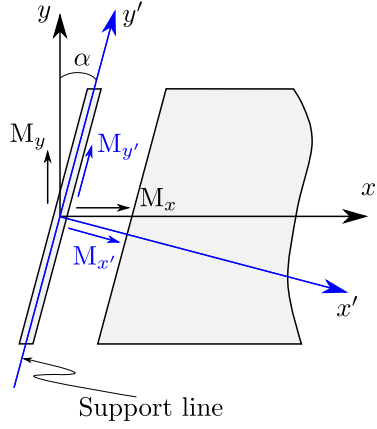}
	\caption{Coordinate systems}
	\label{fig.2}
\end{figure}
\begin{subequations}
	\begin{align}
	\theta_{x'} &= \theta\cos(\alpha) - \frac{\partial u}{\partial x} \sin(\alpha) \\
	M_{y'} &= M_x\sin(\alpha) + M_y\cos(\alpha) = GJ \frac{\partial\theta}{\partial x} \sin(\alpha) + EI \frac{\partial^2 u}{\partial x^2} \cos(\alpha)
\end{align}
\end{subequations}

Hence, the boundary conditions for the problem can be written as:

\begin{subequations}
	\begin{align}
		\phi (0) = \phi (L)=0 \\
		\varphi (0) \cos(\alpha) - \frac{\mathrm{d}}{\mathrm{d} x} \phi (0) \sin(\alpha)= 0 \\
		\varphi (L) \cos(\alpha) - \frac{\mathrm{d}}{\mathrm{d} x} \phi (L) \sin(\alpha)= 0 \\
		GJ \frac{\mathrm{d}}{\mathrm{d} x}\varphi(0) \sin(\alpha) + EI \frac{\mathrm{d}^2}{\mathrm{d} x^2}\phi(0) \cos(\alpha) = 0 \\
		GJ \frac{\mathrm{d}}{\mathrm{d} x}\varphi(L) \sin(\alpha) + EI \frac{\mathrm{d}^2}{\mathrm{d} x^2}\phi(L) \cos(\alpha) = 0 
	\end{align}
\end{subequations}
From these six conditions, a homogeneous system of equations is obtained as:
\begin{equation}
	\mathbf{A}\mathrm{X}  = 0
\end{equation}
where $\mathrm{X}=[C_{n,1},~C_{n,2},~C_{n,3},~C_{n,4},~C_{n,5},~C_{n,6}]^{T}$ is the vector of six constants to be determined, and the matrix $\mathbf{A}$ is expressed as:
\begin{equation}
	\mathbf{A} = \begin{bmatrix}
		0 & 1 & 0 & 1 & 0 & 0 \\
		\sin(\beta L) & \cos(\beta L) & \sinh(\beta L) & \cosh(\beta L) & 0 & 0 \\
		-a_1 & 0 & -a_1 & 0 & 0 & \cos(\alpha) \\
		-a_1\cos(\beta L) & a_1\sin(\beta L) & -a_1\cosh(\beta L) & -a_1\sinh(\beta L) & \sin(\lambda L)\cos(\alpha) & \cos(\lambda L)\cos(\alpha) \\
		0 & -a_2 & 0 & a_2 & a_3 & 0 \\
		-a_2\sin(\beta L) & -a_2\cos(\beta L) & a_2\sinh(\beta L) & a_2\cosh(\beta L) & a_3\cos(\lambda L) & -a_3\sin(\lambda L) 
	\end{bmatrix}
\end{equation}
with $a_1 = \beta\sin(\alpha) $, $a_2 = EI\beta^2\cos(\alpha)$ and $a_3=GJ\lambda\sin(\alpha)$. The eigenvalues are calculated by solving $\det(\mathbf{A}) = 0$. It is noted that the determinant of the matrix $\mathbf{A}$ can be expressed in a function of unique variable $\beta$ ($\lambda = r\beta^2\sqrt{EI/GJ}$). The extraction of the eigenvalues can be performed by using any symbolic mathematical program (e.g. Maple or Matlab). In fact, in this study the symbolic calculation implemented in Matlab is used to extract the values of $\beta$ for desired modes used in the dynamic calculation. The eigenvector corresponding to the $n^{ht}$ mode is obtained by applying singular value decomposition to the matrix $\mathbf{A}$.
\subsection{Orthogonality Relationship}
\label{sec.2.2}
In order to apply the modal superposition technique for solving the forced vibration problems in the skew bridges, it is necessary to determine the orthogonality relationship between the mode shapes. On the basis of the equations \eqref{eq.diff}, these equations can be reformulated by multiplying both sides of these by an arbitrary mode $\phi_m(x)$ and $\varphi_m(x)$, respectively, and integrating with respect to $x$ over the length $L$, one obtains

\begin{subequations}
	\begin{align}
		\int_{0}^{L} \mathrm{EI} \phi_{n}^{''''}(x) \phi_{m}(x)\mathrm{d}x - m\omega_n^2\int_{0}^{L} \phi_{n}(x)\phi_{m}(x) \mathrm{d}x = 0 \label{eq.ref1}\\
		\int_{0}^{L} \mathrm{GJ} \varphi_{n}^{''}(x)\varphi_m(x) + mr^2\omega_n^2\int_0^L \varphi_n(x)\varphi_m(x)\mathrm{d}x = 0 \label{eq.ref2}
	\end{align}
	\label{eq.reformula}
\end{subequations}

By means of using the integration by parts of the left-hand side of the equations \eqref{eq.reformula} (twice for Eq. \eqref{eq.ref1} and once for Eq. \ref{eq.ref2}) and applying the boundary conditions derived for the problem, gives:

\begin{subequations}
	\begin{align}
		\mathrm{EI}\int_{0}^{L} \phi_n^{''}(x)\phi_m^{''}(x) \mathrm{d}x +\mathrm{GJ}\tan(\alpha)[\varphi_n^{'}(L)\phi_m^{'}(L) - \varphi_n^{'}(0)\phi_m^{'}(0)]  - m\omega_n^2\int_{0}^{L} \phi_{n}(x)\phi_{m}(x) \mathrm{d}x &= 0 \\
		\mathrm{GJ} \left[ \tan(\alpha) [\varphi_m^{'}(L)\phi_m^{'}(L) -\varphi_m^{'}(0)\phi_m^{'}(0) ]
  -\int_{0}^{L} \varphi_n^{'}(x) \varphi_m^{'}(x)\mathrm{d}x \right] + mr^2\omega_n^2\int_0^L \varphi_n(x)\varphi_m(x)\mathrm{d}x &= 0
	\end{align}
	\label{eq.reformula3}
\end{subequations}

Interchanging the indices $n$ by $m$ in the equation \eqref{eq.reformula3} and subtracting from its original form, which gives the following relations for any $n\neq m$:

\begin{subequations}
	\begin{align}
		\mathrm{GJ}\tan(\alpha) [\varphi_n^{'}(L)\phi_m^{'}(L) - \varphi_n^{'}(0)\phi_m^{'}(0) - \varphi_m^{'}(L)\phi_n^{'}(L) + \varphi_m^{'}(0)\phi_n^{'}(0) ] &  \nonumber \\ - m(\omega_n^2-\omega_m^2)\int_{0}^{L} \phi_{n}(x)\phi_{m}(x) \mathrm{d}x =  0 &  \label{eq.reformula41} \\
		\mathrm{GJ}\tan(\alpha) [\varphi_n^{'}(L)\phi_m^{'}(L) - \varphi_n^{'}(0)\phi_m^{'}(0) - \varphi_m^{'}(L)\phi_n^{'}(L) + \varphi_m^{'}(0)\phi_n^{'}(0) ] &  \nonumber \\  mr^{2}(\omega_n^2-\omega_m^2)\int_{0}^{L} \varphi_{n}(x)\varphi_{m}(x) \mathrm{d}x = 0 &  \label{eq.reformula42}
	\end{align}
\end{subequations}
Next,  subtracting the equation \eqref{eq.reformula41} from the equation \eqref{eq.reformula42} gives rise to:

\begin{equation}
	(\omega_n^2-\omega_m^{2})\left(m\int_{0}^{L} \phi_{n}(x)\phi_{m}(x) \mathrm{d}x + mr^2\int_{0}^{L} \varphi_{n}(x)\varphi_{m}(x) \mathrm{d}x  \right) = 0
	\label{eq.reformula5}
\end{equation}
Due to the fact that $\omega_n \neq \omega_m$ and $\int_{0}^{L} \phi_{n}(x)\phi_{m}(x) \mathrm{d}x \geq 0$ and $\int_{0}^{L} \varphi_{n}(x)\varphi_{m}(x) \mathrm{d}x \geq 0$ 
for any $n \neq m$, the condition established in Eq. \eqref{eq.reformula5} will be fulfilled when:

\begin{subequations}
	\begin{align}
		m\int_{0}^{L} \phi_{n}(x)\phi_{m}(x) \mathrm{d}x = 0  \\
		mr^2\int_{0}^{L} \varphi_{n}(x)\varphi_{m}(x) \mathrm{d}x  = 0
	\end{align}
\end{subequations}
which corresponds to the orthogonality relationship of the simply-supported skew bridge.

\subsection{Vibration induced by a moving load and a convoy of moving loads}
\label{sec.2.3}
Once the natural frequencies and the associated mode shapes are found, and the orthogonality relationship between the modes is known, it is possible to apply the modal superposition technique for obtaining the response of the skew bridge due to a moving load. The vertical load and the twisting moment apply on the bridge deck can be determined as:

\begin{subequations}
\begin{align}
	p(x,t) &= P\delta(x-vt) \\
	m_t(x,t) &= \frac{P\delta(x-vt)L(\epsilon-\epsilon^2)\cot(\alpha)}{2(1+K\cot^2(\alpha))} + P\delta(x-vt)e \label{eq.mov_load2}
\end{align}
\label{eq.mov_load}
\end{subequations}
where $P$ is the magnitude of the moving load, $\delta$ is the Dirac delta function, $\epsilon = vt/L$, $K = EI/GJ$ and $e$  is the load eccentricity respect to the mass centre of the bridge deck section. The first part of the right side of Eq. \eqref{eq.mov_load2} is due to the skewness of the bridge \cite{Kollbrunner1969, Manterola2006}, and the second part is due to the load eccentricity. Using the modal superposition technique and applying the orthogonality relationship, the differential equations in the generalized coordinates are uncoupled: 

\begin{subequations}
	\begin{align}
		\ddot{q}_n(t) + 2\xi\omega_n \dot{q}_n(t) + \omega_n^2 q_n(t) &= \frac{P\phi_n(vt)}{\int_{0}^{L} m \phi_n(x)^2 \mathrm{d}x }\\
		\ddot{p}_n(t) + \omega_n^2 p_n(t) &= \frac{1}{\int_{0}^{L} m r^2 \varphi_n(x)^2 \mathrm{d}x } \left [\frac{P L (\epsilon-\epsilon^2)\cot(\alpha)}{2(1+K\cot^2(\alpha))} + P e \right]\varphi_n(vt)
	\end{align}
	\label{eq.transfor3}
\end{subequations}
In order to solve the differential equations \eqref{eq.transfor3}, several techniques can be applied. In this work, the solution of Eq. \eqref{eq.transfor3} is obtained by using the integration method based on the interpolation of excitation \cite{Chopra2012}, which has the advantage that it gives an exact solution and a highly efficient numerical procedure. The solution of Eq. \eqref{eq.transfor3} at time $i+1$ can be determined as:
\begin{equation}
	\mathrm{w}_{i+1} = A \mathrm{w}_i + B \dot{\mathrm{w}}_i + C Q_i + D Q_{i+1} \\
\end{equation}
and its velocity is given by

\begin{equation}
	\dot{\mathrm{w}}_{i+1} = A' \mathrm{w}_i + B' \dot{\mathrm{w}}_i + C' Q_i + D' Q_{i+1} \\
\end{equation}
where $\mathrm{w}=[q_n,p_n]^T$, $Q = [\frac{P\phi_n(vt)}{\int_{0}^{L} m \phi_n(x)^2 \mathrm{d}x }, ~ \frac{1}{\int_{0}^{L} m r^2 \varphi_n(x)^2 \mathrm{d}x } \left [\frac{P L (\epsilon-\epsilon^2)\cot(\alpha)}{2(1+K\cot^2(\alpha))} + P e \right]\varphi_n(vt)]^T$, and $A,~B,~C,~...,~D'$ are the coefficients that depend on the structure parameters $\omega_n,~\xi_n$ and on the time step $\Delta t$ (detail formulations can be found in Appendix A)

\begin{figure}[h!]
	\centering
	\subfigure[a moving load]{\label{fig.31}\includegraphics[width=0.35\textwidth]{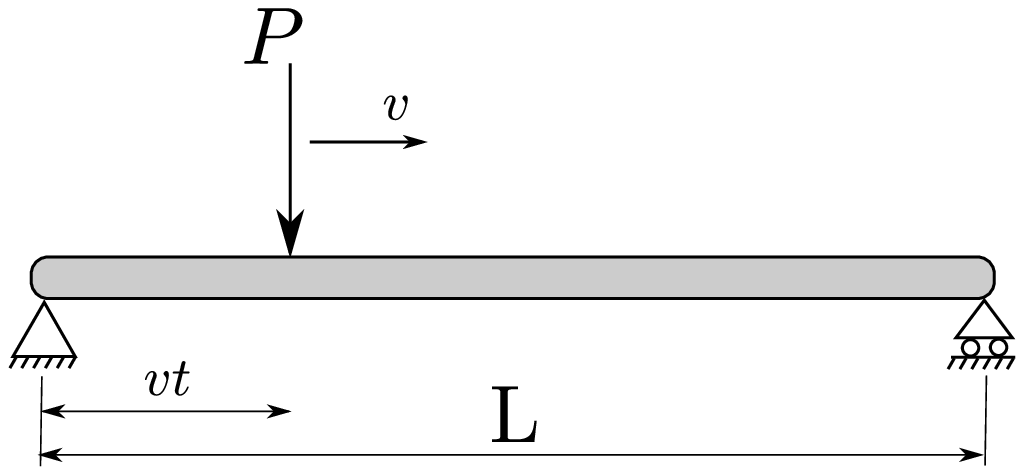}}
	\subfigure[a convoy of moving loads]{\label{fig.32}\includegraphics[width=0.55\textwidth]{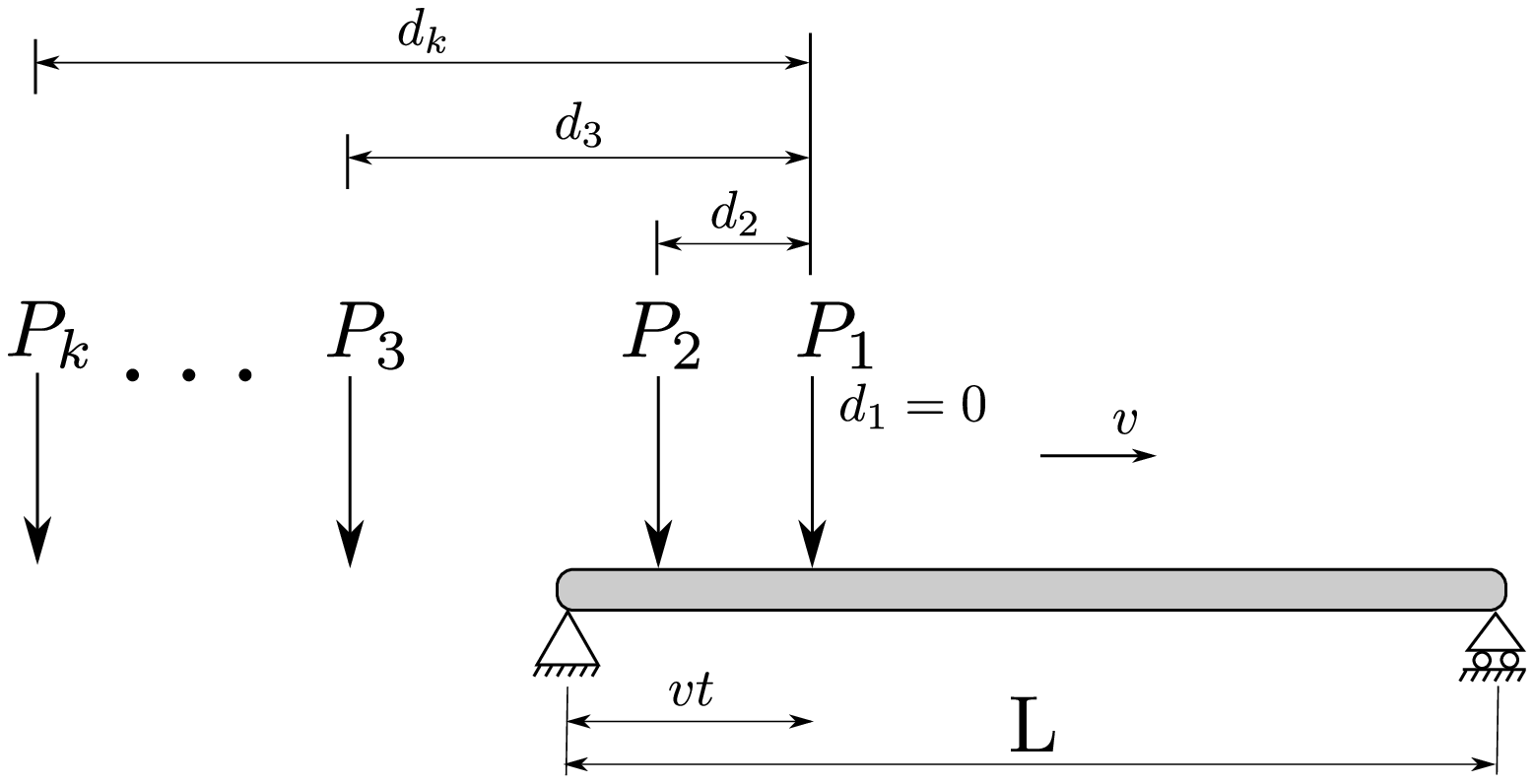}}
	\caption{Moving loads}
	\label{fig.3}
\end{figure}

For the case that the bridge is forced by a convoy of moving loads as shown in Fig. \ref{fig.32}, the uncoupled differential equations in the generalized coordinates for each mode of vibration $n$ are given as

\begin{subequations}
	\begin{align}
		\ddot{q}_n(t) + 2\xi\omega_n \dot{q}_n(t) + \omega_n^2 q_n(t) &=\sum_{k=1}^{n_{P}} \frac{P_k\phi_n(v t - d_k)}{\int_{0}^{L} m \phi_n(x)^2 \mathrm{d}x }\\
		\ddot{p}_n(t) + \omega_n^2 p_n(t) &= \sum_{k=1}^{n_{P}} \frac{\varphi_n(vt - d_k)}{\int_{0}^{L} m r^2 \varphi_n^2(x) \mathrm{d}x }\left [\frac{P_k L (\epsilon_k-\epsilon_k^2)\cot(\alpha)}{2(1+K\cot^2(\alpha))} + P_k e \right]
	\end{align}
	\label{eq.transfor4}
\end{subequations}
where $n_P$ is the number of moving loads, $d_k$ is the distance between the first load and the $k^{th}$ load, $P_k$ is the magnitude of the $k^{th}$ load, and $\epsilon_k = (vt-d_k)/L$. The solution of Eq. \eqref{eq.transfor4} is obtained in similar way as in the case of a moving load. Attention needs to be paid in the determination of the modal loads in the right side of Eq. \eqref{eq.transfor4}. For the loads that do not enter the bridge ($vt-d_k < 0$) or leave the bridge ($vt-d_k > L$) the modal loads associated with those loads are zero.

\section{A simplified model}
\begin{figure}
	\centering
	\includegraphics[width=\textwidth]{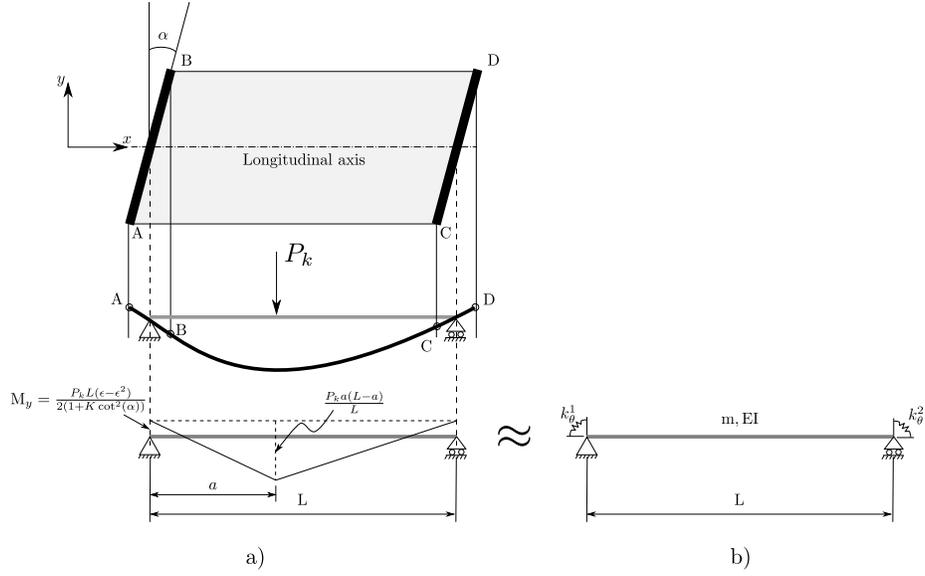}
	\caption{a) Diagram of bending moment of a skew bridge under a static load, b) Simplified model adopted for simply-supported skew bridge}
	\label{fig.4}
\end{figure}

In this part of the work, a simplified 2D model is developed in order to assimilate the effect of the skewness of the support on the vertical vibration of the simply-supported skew bridges. It is well known that the skewness of the supports causes the torsional moment on the bridge even for the vertical, centric loads. Those torsional moments in turn have a certain influence on the bending moment. In particular, a negative bending moment is introduced at the supports as shown in Fig. \ref{fig.4}a \cite{Kollbrunner1969, Manterola2006}, making that for the purpose of vertical flexure the simply-supported skew beam behaves like as an elastically-fixed beam, or in other words, as a beam with rotational support with stiffness $k_{\theta}$ as shown in Fig. \ref{fig.4}b. It is noted that the negative bending moments at the supports change with the load position on the bridge. Therefore, the stiffness of the rotational support is also changed and can be different at different supports. In order to simplify the calculation the stiffness of the rotational support are considered the same in both supports. With this assumption, the stiffness of the rotational support can be determined as:

\begin{equation}
	k_{\theta}^1 = k_{\theta}^2 = k_{\theta} = \frac{2\mathrm{GJ}}{L\cot^2(\alpha)}
\end{equation}

In additions to the previously adopted assumptions, the following additional assumptions are used for the simplified model:
\begin{itemize}
	\item Only the vertical vibration is taken into account in the model. The load eccentricity is not considered.
	\item The bridge deck is modelled by the 2D Euler-Bernoulli beam theory.
\end{itemize}
\subsection{Natural frequencies and mode shapes}
The governing equation for the free vibration of the simplified model is similar to Eq. \eqref{eq.diff1}. The solution of this equation is given in \eqref{eq.mode_shape1}. The determination of the frequencies and its correspondent mode shapes is solving the homogeneous system of equations:

\begin{equation}
	\mathbf{B}\mathrm{J} = 0
\end{equation}
where $\mathrm{J} = [C_{1},C_{2},C_{3},C_4]^T$ is a vector containing the four mode shape coefficients, $\mathbf{B}$ is the characteristic matrix that can be determined by applying the boundary  conditions. For the simplified model proposed in this study, the boundary conditions are:
\begin{itemize}
	\item There is not vertical displacement at the supports
		\begin{equation}
			u(0,t)=u(L,t)=0 \implies \phi_n(0) = \phi_n(L)= 0 
		\end{equation}
	\item Equilibrium of moments at the supports:
		\begin{align}
			EI \left.\frac{\partial^2 u}{\partial x^2} \right|_0 &= k_{\theta} \left.\frac{\partial u}{\partial x}\right|_0 && \implies EI\phi''_n(0) = k_{\theta}\phi'_n(0)\\
			EI \left.\frac{\partial^2 u}{\partial x^2} \right|_L &= -k_{\theta} \left.\frac{\partial u}{\partial x}\right|_L && \implies EI\phi''_n(L)=-k_{\theta}\phi'_n(L) 
		\end{align}
\end{itemize}
Therefore, the characteristic matrix $\mathbf{B}$ is obtained as

\begin{equation}
	\mathbf{B} = \begin{bmatrix}
		0 & 1 & 0 & 1 \\
		\sin(\beta L) & \cos(\beta L) & \sinh(\beta L) & \cosh(\beta L) \\
		-k_{\theta}\beta & -EI\beta^2 & -k_{\theta}\beta & EI\beta^2 \\
		a_{41} & a_{42} & a_{43} & a_{44}
	\end{bmatrix}
\end{equation}
in which 
\begin{align}
	a_{41} &= k_{\theta}\beta \cos(\beta L) - EI\beta^2\sin(\beta L) ;& a_{42} &= -k_{\theta}\beta \sin(\beta L) - EI\beta^2\cos(\beta L); \\
	a_{43} &= k_{\theta}\beta \cosh(\beta L) + EI\beta^2\sinh(\beta L) ;& a_{44} &= k_{\theta}\beta \sinh(\beta L) + EI\beta^2\cosh(\beta L) 
\end{align}
The procedure to obtain the eigenvalues and eigenvector is similar to the previously described in section \ref{sec.2.1}. 
\subsection{Orthogonality relationship}
Similar to the analysis in section \ref{sec.2.2}, the equation \eqref{eq.diff1} can be rewritten, using the boundary conditions of the simplified model, as:

\begin{equation}
	\mathrm{EI}\int_{0}^{L} \phi''_n(x)\phi''_m(x)\mathrm{d}x +k_{\theta}[\phi'_n(0)\phi'_m(0)+\phi'_n(L)\phi'_m(L)] -m\omega^2_n\int_{0}^{L} \phi_n(x)\phi_m(x)\mathrm{d}x = 0
	\label{eq.orth_simpl_model}
\end{equation}
Interchanging the indices $n$ and $m$ in Eq. \eqref{eq.orth_simpl_model} and subtracting the resulting equation from its original form gives
\begin{equation}
	m(\omega^2_n-\omega^2_m)\int_{0}^{L} \phi_n(x)\phi_m(x)\mathrm{d}x = 0 ~ \text{or}~ m\int_{0}^{L} \phi_n(x)\phi_m(x)\mathrm{d}x = 0 ~ (\omega_n \neq \omega_m)
\end{equation}
which is the orthogonality relationship between the mode shapes for the simplified model.
\subsection{Vibration-induced by a moving load and a convoy of moving loads}
The dynamic response of the bridge under moving loads is obtained by using the same way described for the analytical model in section \ref{sec.2.3}. The only difference is that the torsional response is eliminated in the calculation. 
\section{Numerical validations}
Two numerical examples are used in order to validate the proposed models. The results obtained by proposed models are compared with those obtained by Finite Element (FE) simulations. For each example, a FE model is developed in the program FEAP \cite{feap}, built with 3D Euler-Bernoulli beam element (stick model). A moving load or a convoy of moving loads is applied to the nodal forces along the centreline axis, using time-dependent amplitude functions. The dynamic responses in FE models are obtained by solving in the time domain using the modal superposition technique with a time step of $0.001$ s. For all examples, the first five modes of vibration  are considered in the calculation and  a constant damping ratio is assumed for all considered modes ($\xi_n = \xi$). Attention should be paid to select the total number of modes of vibration considered in the FE models, since the first five modes of vibration obtained by FE model are not always corresponding to the first five modes obtained by analytical and simplified models. 

\begin{figure}[h!]
	\centering
	\subfigure[]{\label{fig.slab}\includegraphics[width=0.48\textwidth]{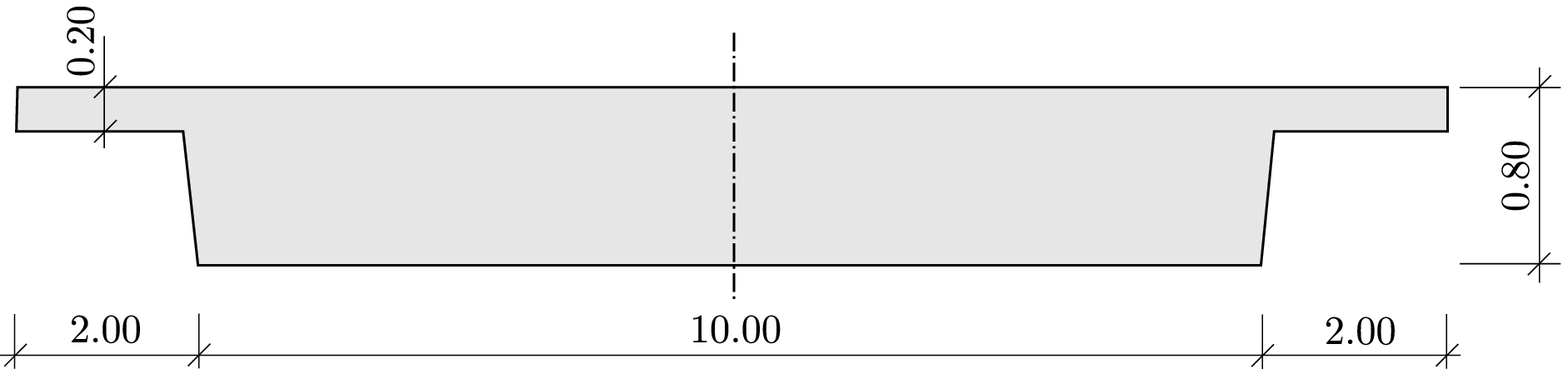}}
	\subfigure[]{\label{fig.box_slab}\includegraphics[width=0.48\textwidth]{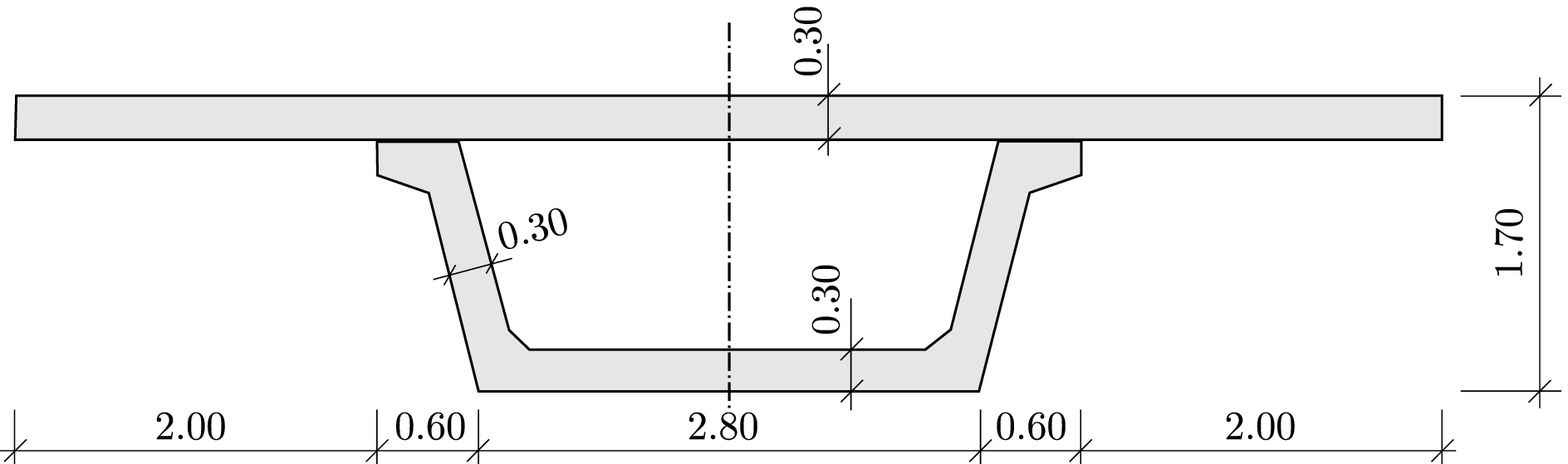}}
	\caption{Cross sections: a) for example 1, b) for example 2}
\end{figure}
\subsection{Example 1: a simply-supported skew slab bridge under a moving load}
\begin{figure}[h!]
	\centering
	\subfigure[]{\label{}\includegraphics[width=0.8\textwidth]{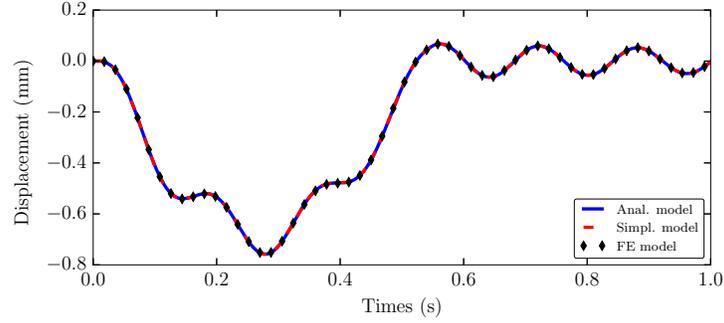}}
	\subfigure[]{\label{}\includegraphics[width=0.8\textwidth]{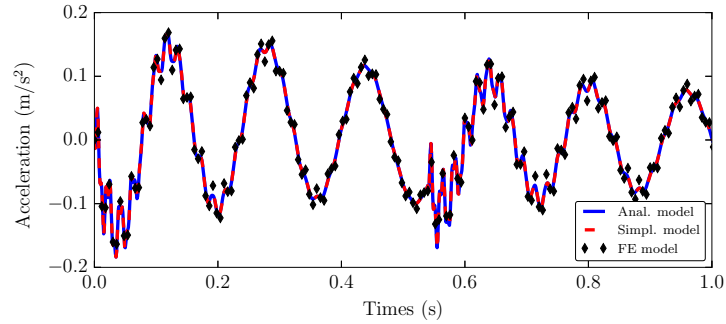}}
	\caption{Dynamic responses at the mid-span under a moving load: a) displacement, b) acceleration}
	\label{fig.slab_bridge}
\end{figure}

A simply-supported skew slab bridge is considered in this example. The skew angle of the bridge is $20^{o}$. The bridge clear-span is 15.0 m. The cross section of the bridge is shown in Fig. \ref{fig.slab} and the following geometric and mechanical characteristics are used in the calculation:
\begin{itemize}
	\item Elastic modulus $E$ is $3.2e10 $ N/m$^2$ with Poisson coefficient $\nu=0.25$.
	\item Properties of the cross section: $I = 0.4987$ m$^4$, $J=1.7067$ m$^4$, $m=22.5$ t/m and $r=0.2354$ m.
	\item Damping ratio $\xi$ is 2\%.
\end{itemize}
The bridge is subjected to the action of a moving load of 170 kN with a constant speed of 100 km/h. The frequencies of the first five modes considered in the calculation are extracted and listed in Table \ref{table.1} for all models. It can be noted that there is a very good agreement in the natural frequency between the analytical, simplified and FE models. In fact, the maximum difference in the frequency between models does not exceed 2\%. The similar agreement is also observed with the dynamic responses in terms of vertical displacement and acceleration at the mid-span for three models, as shown in Fig. \ref{fig.slab_bridge}. From this result it can be remarked that the proposed simplified model is enable to simulate the vertical dynamic response of the simply-supported skew bridge.
\begin{table}[h!]
	\centering
	\caption{Frequencies of first five modes of vibration of different models (in Hz)}
	\begin{tabular}{ccccc} \hline
		Modes & Anal. Model & Simpl. Model & FE Model & Description \\ \hline
		1     & 6.259 & 6.259 & 6.259 & 1$^{st}$ mode in FE model\\ 
		2     & 23.548 & 23.910 & 23.518 & 2$^{nd}$ mode in FE model\\
		3     & 53.361 & 53.312 & 53.312 & 3$^{rd}$ mode in FE model\\
		4     & 94.423 & 94.471 & 94.072 & 4$^{th}$ mode in FE model\\
		5     & 147.716 & 147.387 & 147.387 & 6$^{th}$ mode in FE model  \\ \hline
	\end{tabular}
	\label{table.1}
\end{table}

\subsection{Example 2: a simply-supported skew box-slab bridge under a convoy of moving loads}
This example attempts to simulate the dynamic response of a railway bridge under an HSLM A1 train \cite{CEN2003} which is the desired application of the proposed analytical and simplified methods presented in this paper. The studied bridge is a typical box-slab bridge designed for single-track and has a cross section as shown in Fig. \ref{fig.box_slab}. A skew angle of 10$^{o}$ is considered. The bridge clear-span is 24.0 m. The geometric and mechanical properties of the bridge's cross section used in the calculation are:
\begin{itemize}
	\item Elastic modulus $E = 3.2e10$ n/m$^2$ with Poisson coefficient of 0.25.
	\item $I = 1.3921$ m$^4$, $J=2.6741$ m$^4$, $m=9.774$ t/m and $r=0.5967$ m.
	\item Damping ratio $\xi$ is 1\%.
\end{itemize}
The HSLM-A1 train consists of 18 intermediate coaches, a power coach and a end coach on either sides of the train. In total, the train has 50 axles with a load of 170 kN/axle. The dynamic analysis are carried out for different train speeds ranging from 100 km/h to 300 km/h in increment of 5 km/h. The vertical displacement and acceleration at the mid-span are obtained and compared between the models. The envelope of maximum vertical displacement and acceleration are also depicted for all models in order to validate the proposed analytical and simplified model presented in this paper.

\begin{figure}[h!]
	\centering
	\subfigure[]{\label{}\includegraphics[width=0.85\textwidth]{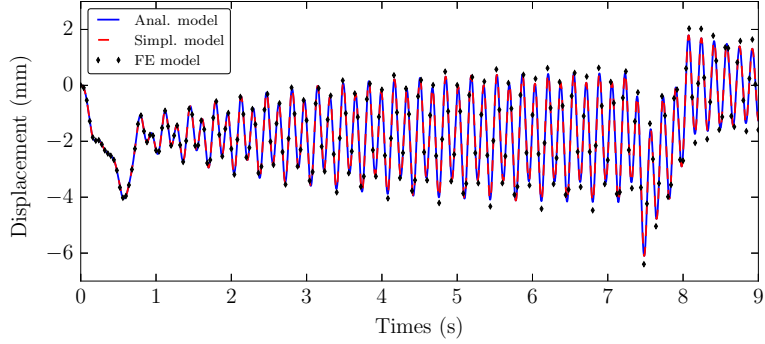}}
	\subfigure[]{\label{}\includegraphics[width=0.85\textwidth]{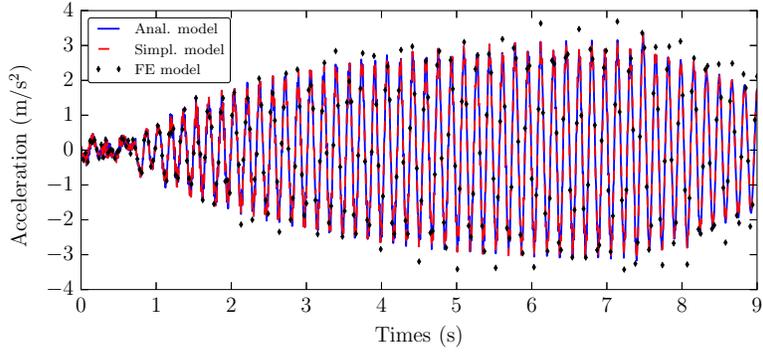}}
	\caption{Dynamic responses at the mid-span under the HSLM-A1 train at velocity of 190 km/h: a) displacement, b) acceleration}
	\label{fig.viga_resonance}
\end{figure}

\begin{figure}[h!]
	\centering
	\subfigure[]{\label{}\includegraphics[width=0.85\textwidth]{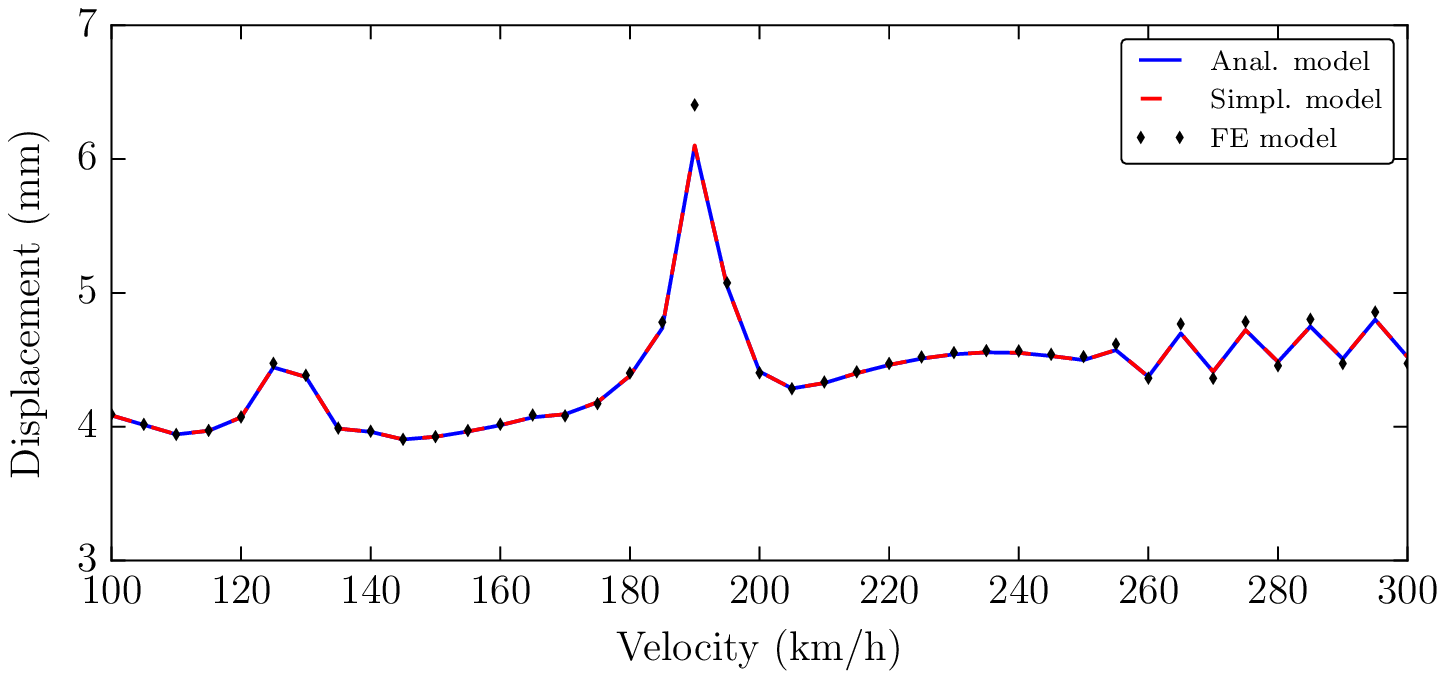}}
	\subfigure[]{\label{}\includegraphics[width=0.85\textwidth]{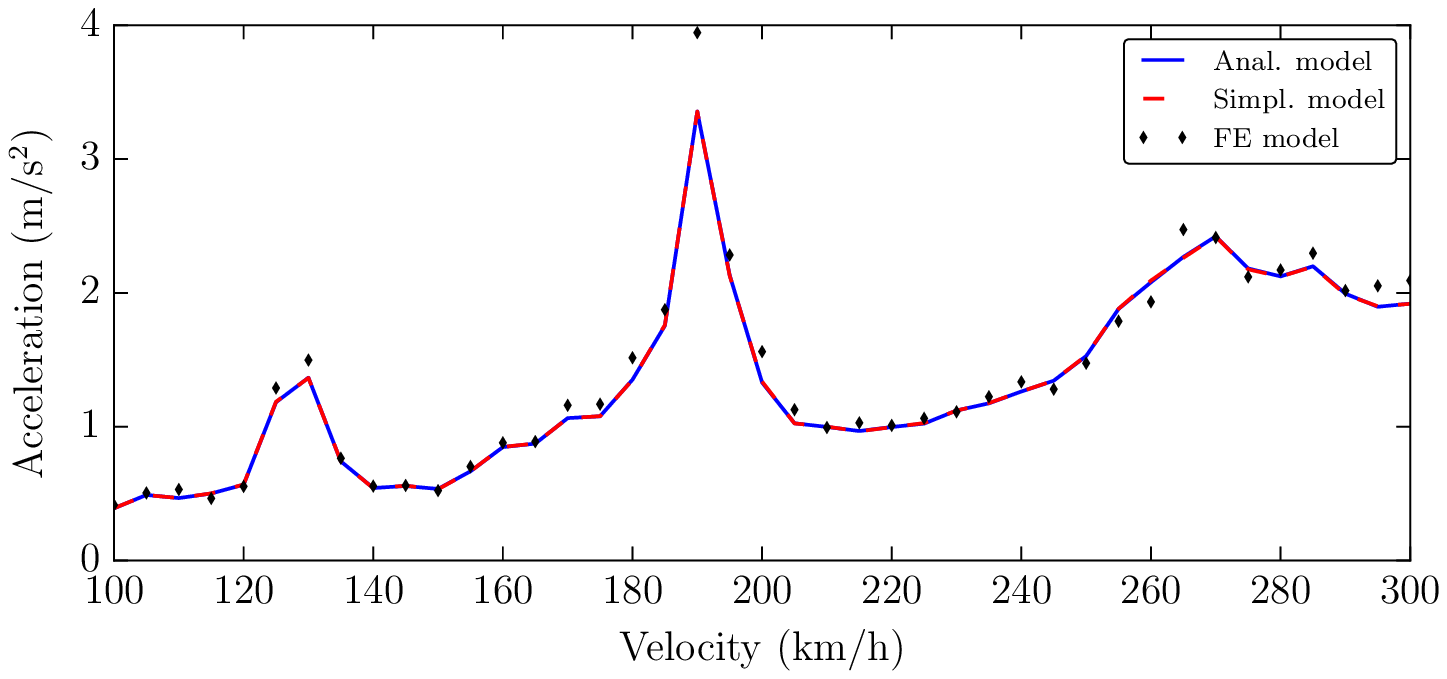}}
	\caption{Envelope of the maximum response at the mid-span under the HSLM-A1 train: a) displacement, b) acceleration}
	\label{fig.envelope}
\end{figure}

Table \ref{table.2} gives the natural frequencies of the first five modes of vibration considered in the calculation. It is known that for the simply-supported bridge the train velocities of resonance can be estimated using the following formula \cite{CEN2003}:

\begin{equation}
	v_i = f_0\frac{D}{i} ~\text{with} ~ i=1,2,3,...,\infty
	\label{eq.velo}
\end{equation}
where $f_0$ is the fundamental frequency; $D$ is the regular distance between load axles and is 18 m for the HSLM-A1 train. According with this Eq. \eqref{eq.velo}, the first three resonance peaks occur at train velocities of almost 382 km/h, 191 km/h and 127 km/h. The dynamic response at the train speed of 190 km/h is shown in Fig. \ref{fig.viga_resonance}. It can be observed that at this train speed (near the second critical speed) the responses are amplified by each axle passing the bridge. The envelope curves for the maximum vertical displacement and acceleration at the mid-span are shown in Fig. \ref{fig.envelope}. It can be noted in Fig. \ref{fig.envelope} that in the considered range of train velocities two peaks of response (both displacement and acceleration) occur at speeds of 190 km/h and 125 km/h which are closed to the predicted critical trains. Therefore, it can be remarked that the estimation of the train velocities of resonance proposed by \cite{CEN2003} is still valid for the skew bridge. Furthermore, from both Figs. \ref{fig.viga_resonance}  and \ref{fig.envelope} it can be concluded that the results obtained using the analytical and simplified model agree well with the ones obtained using the FE model. It should be noted that the time consumed for the calculation using the analytical or simplified model is approximately 50 times faster than the ones using the FE model: the CPU time required for completing a analysis using the analytical model was 1.9 s while 130.5 s was the time for FE model in a standard PC equipped with Intel Xeon processor of 2.33 GHz and 4 GB of RAM.   

\begin{table}[h!]
	\centering
	\caption{Frequencies of first five modes of vibration of different models (in Hz)}
	\begin{tabular}{ccccc} \hline
		Modes & Anal. Model & Simpl. Model & FE Model & Description \\ \hline
		1     & 5.878 & 5.878 & 5.878 & 1$^{st}$ mode in FE model\\
		2     & 23.305 & 23.344 & 23.288 & 2$^{nd}$ mode in FE model\\
		3     & 52.482 & 52.455 & 52.455 & 4$^{th}$ mode in FE model\\
		4     & 93.276 & 93.209 & 93.153 & 6$^{th}$ mode in FE model\\
		5     & 145.748 & 145.608 & 145.642 & 8$^{th}$ mode in FE model\\ \hline
	\end{tabular}
	\label{table.2}
\end{table}
\section{Parametric study}
In this part of the paper, three parametric studies are performed using the simplified model in order to identify parameters that influence significantly the vertical dynamic response of the simply-supported skew bridge under the moving loads. In each study, the value of the studied parameter are changed. The dynamic responses under the HSLM-A1 train corresponding to each value of the parameter are obtained and depicted in function of the studied parameter. The basic properties of the skew bridge in Example 2 are adopted in this section. 
\subsection{Effect of skew angle}
\begin{figure}[h!]
	\centering
	\subfigure[]{\label{fig.influence_skewness1}\includegraphics[width=0.48\textwidth]{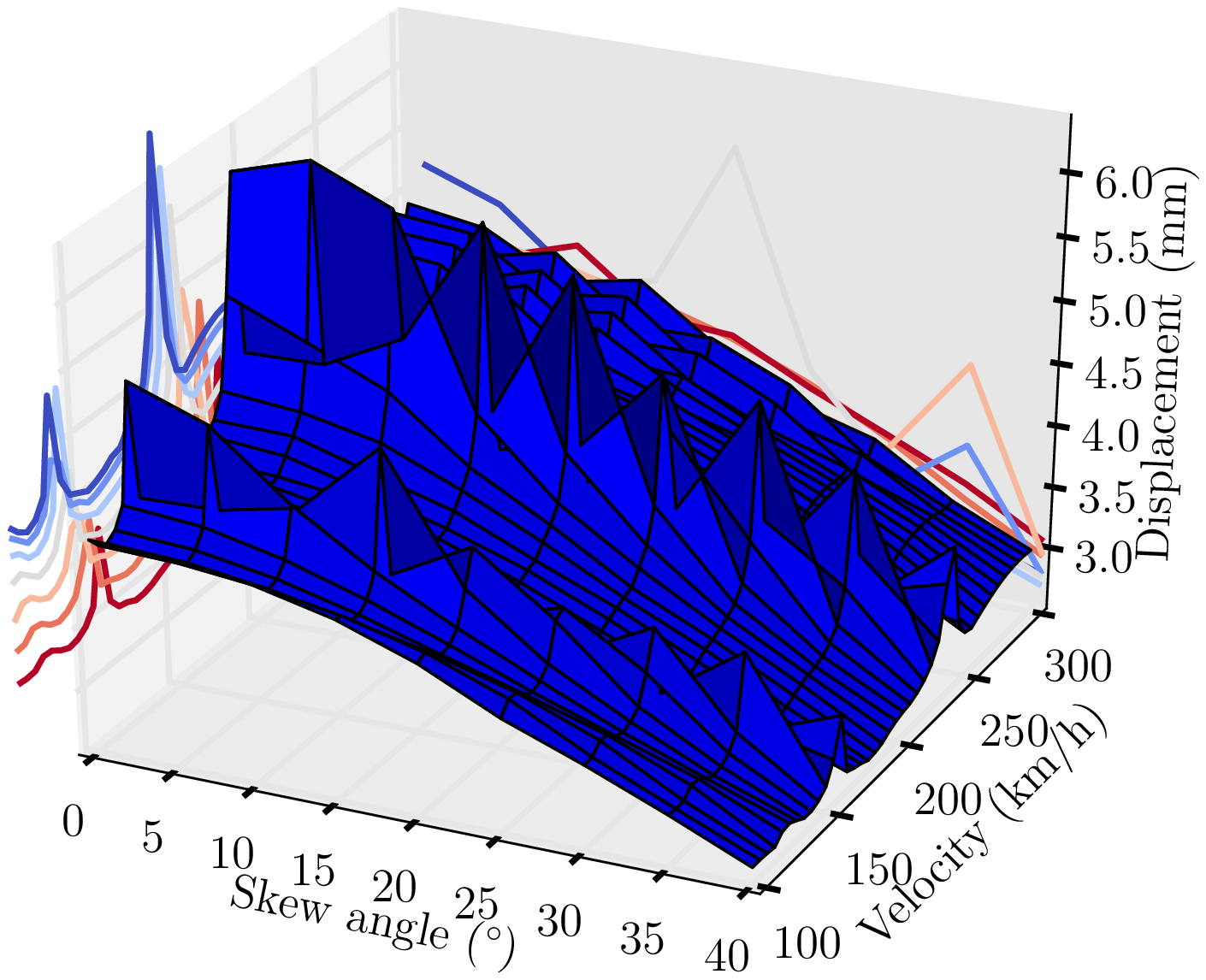}}
	\subfigure[]{\label{fig.influence_skewness2}\includegraphics[width=0.48\textwidth]{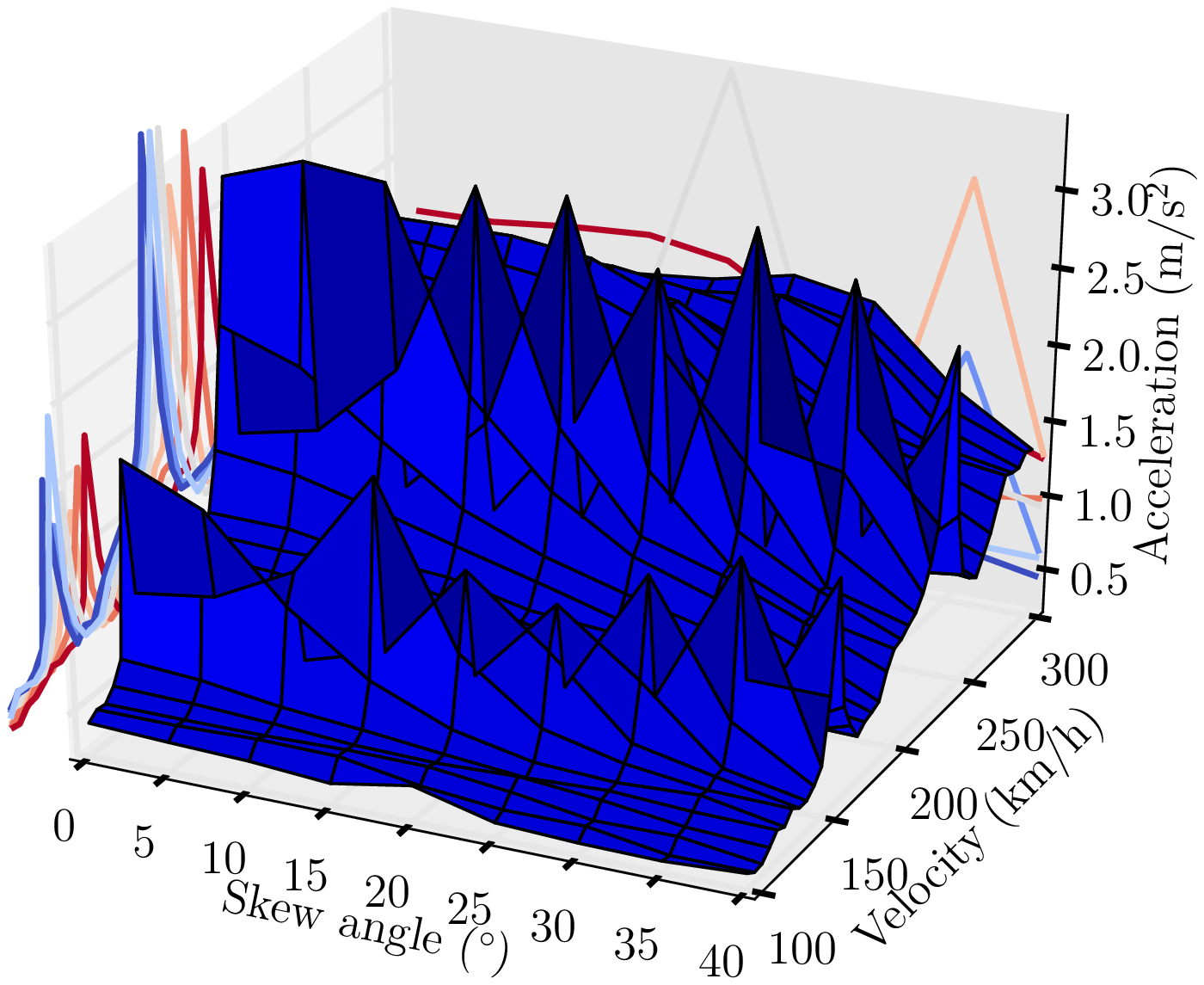}}
	\caption{Effect of  skewness on the dynamic responses: a) displacement, b) acceleration}
	\label{fig.influence_skewness}
\end{figure}

Figure \ref{fig.influence_skewness} shows how the maximum dynamic responses vary with the skew angle when the bridge is forced by the HSLM-A1 train. It can be observed from Fig. \ref{fig.influence_skewness} that the skewness has an important influence on the maximum vertical displacement at the mid-span of the bridge: in general the displacement decreases as the skew angle increases. A sharp change in slope can be observed at the skew angle of 15$^{\circ}$. From this value of the skew angle, the displacement decreases more quickly. Furthermore, the changing in the train velocity of resonance is also observed when the skewness is changed. In fact, the train velocity of resonance increases as the skewness increases. Regarding the maximum acceleration at the mid-span, the skew angle does not has pronounced influence on it: the acceleration hardly increases when the skew angle grows.

\subsection{Effect of torsional to flexural stiffness ratio}
\begin{figure}[h!]
	\centering
	\subfigure[]{\label{fig.influence_GJ_1}\includegraphics[width=0.48\textwidth]{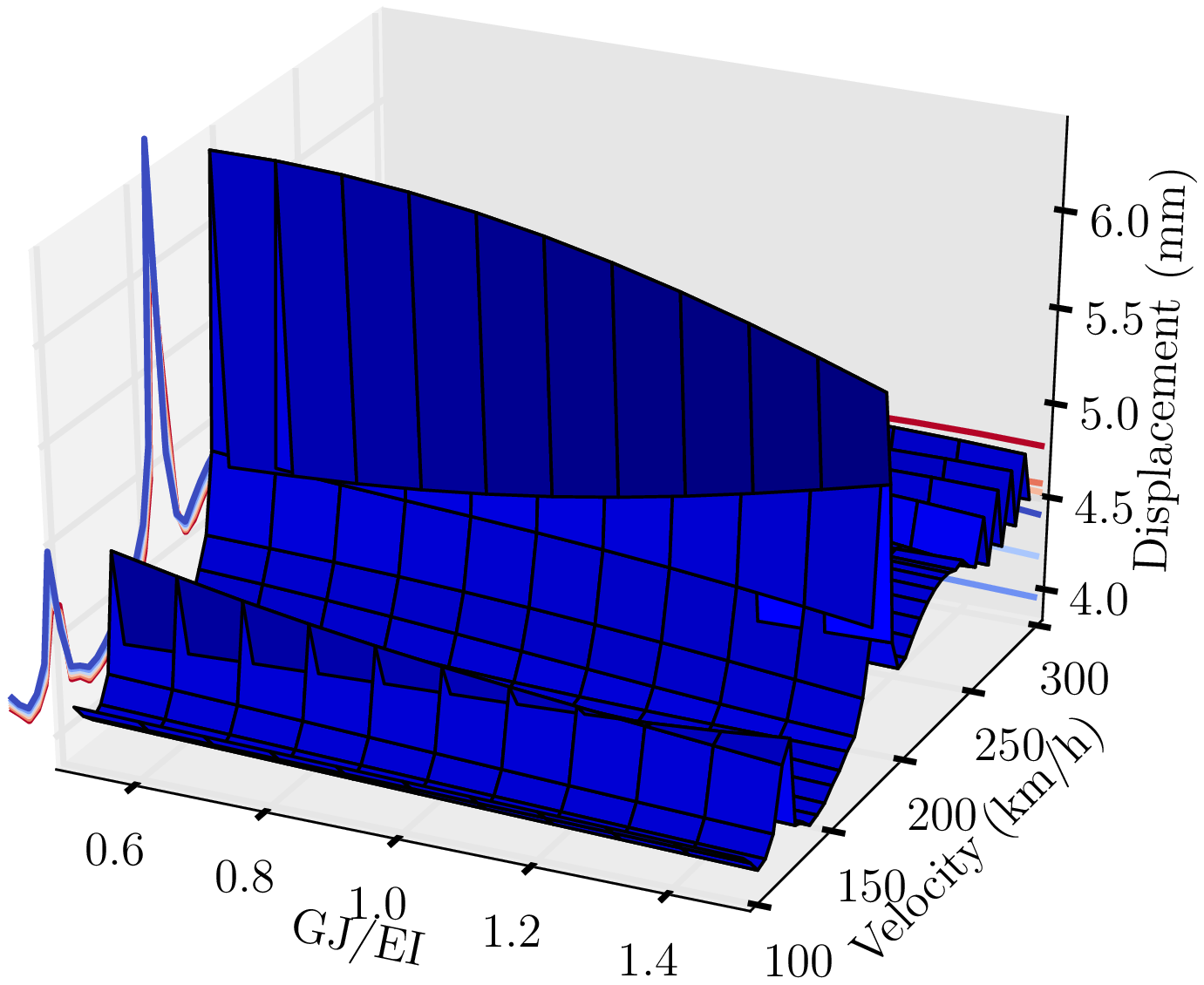}}
	\subfigure[]{\label{fig.influence_GJ_2}\includegraphics[width=0.48\textwidth]{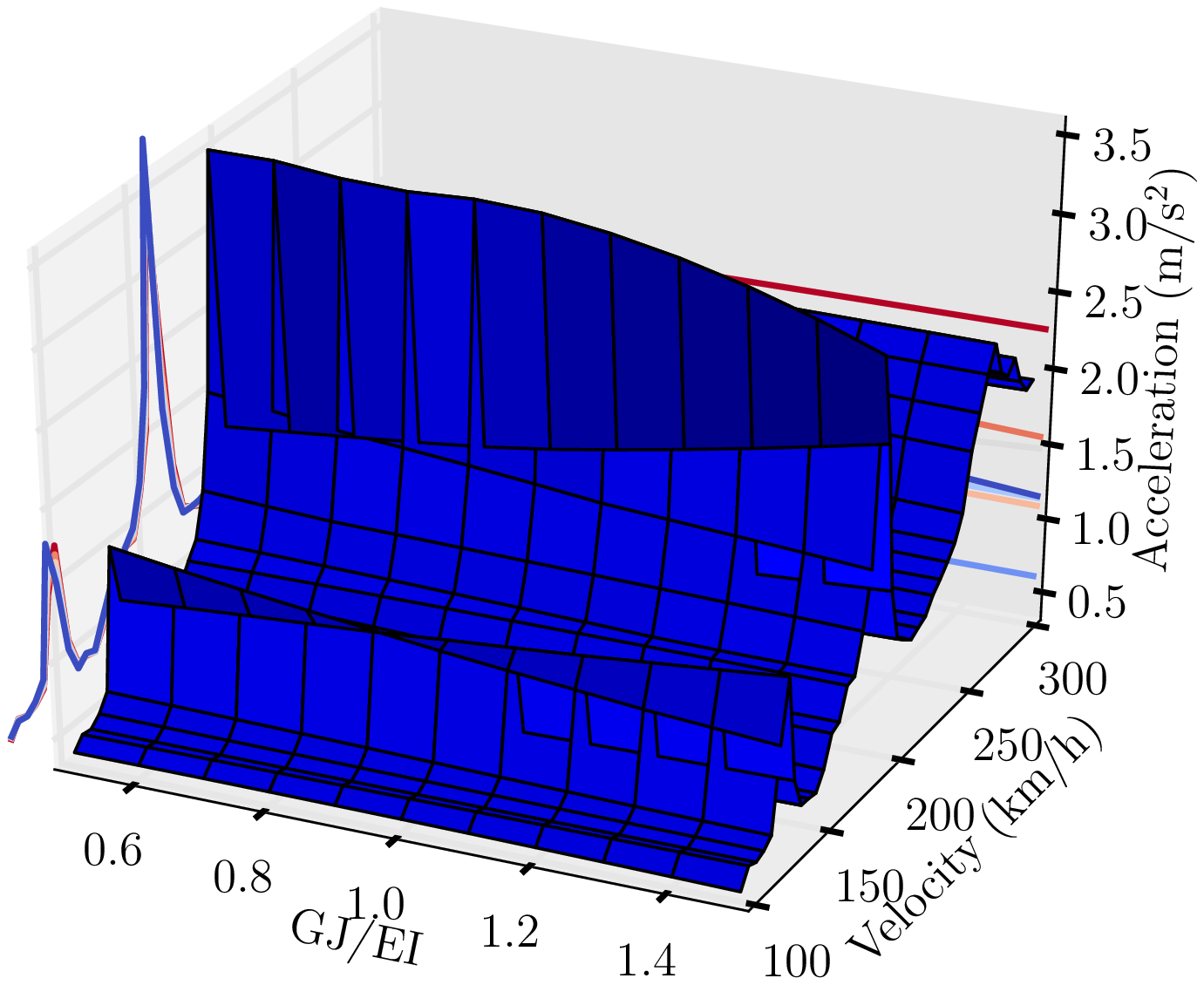}}
	\caption{Effect of torsional to flexural stiffness ratio on the dynamic responses for a skew angle of 10$^{\circ}$: a) displacement, b) acceleration}
	\label{fig.influence_GJ}
\end{figure}

\begin{figure}[h!]
	\centering
	\subfigure[]{\label{fig.influence_GJ_skew30_1}\includegraphics[width=0.48\textwidth]{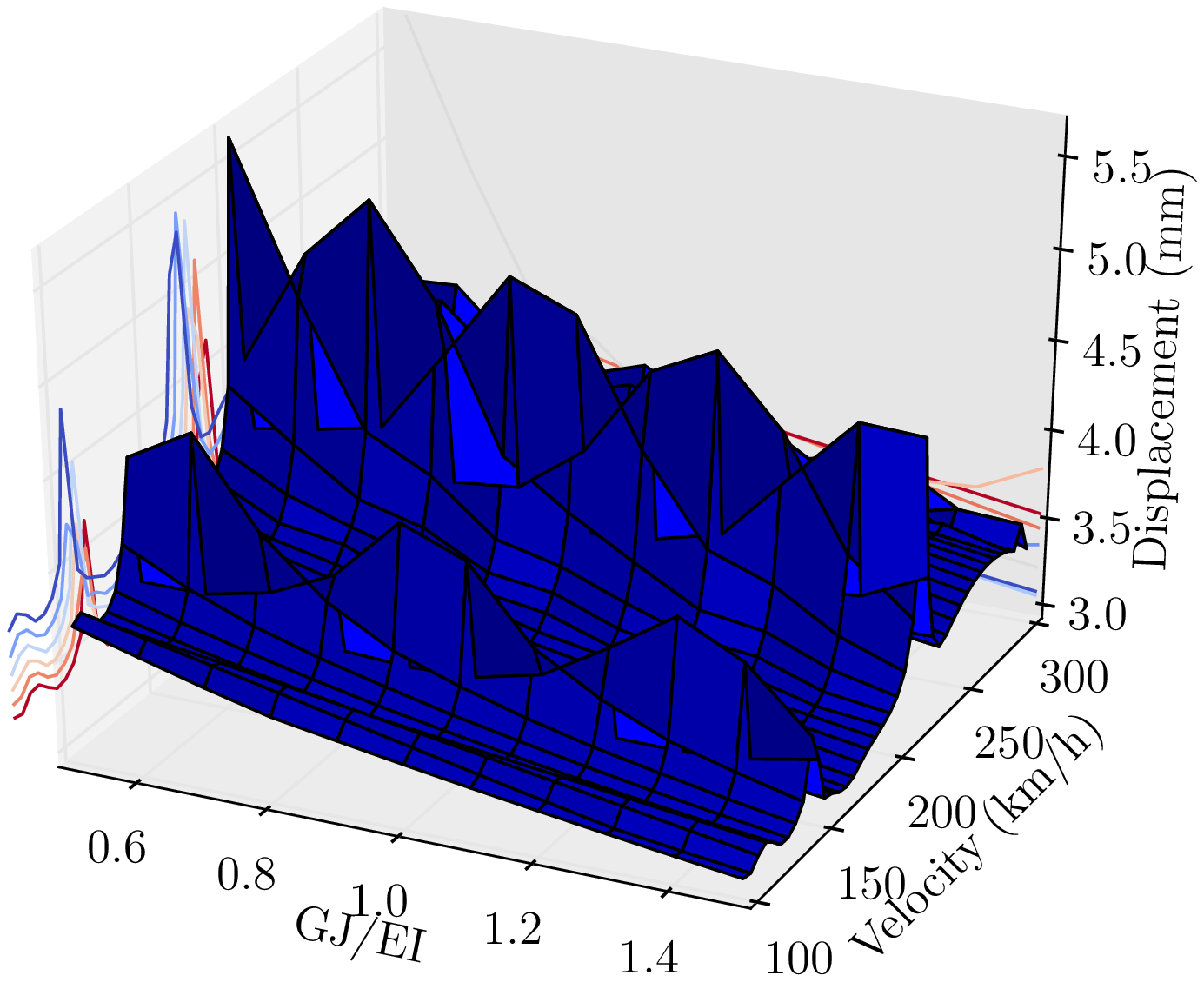}}
	\subfigure[]{\label{fig.influence_GJ_skew30_2}\includegraphics[width=0.48\textwidth]{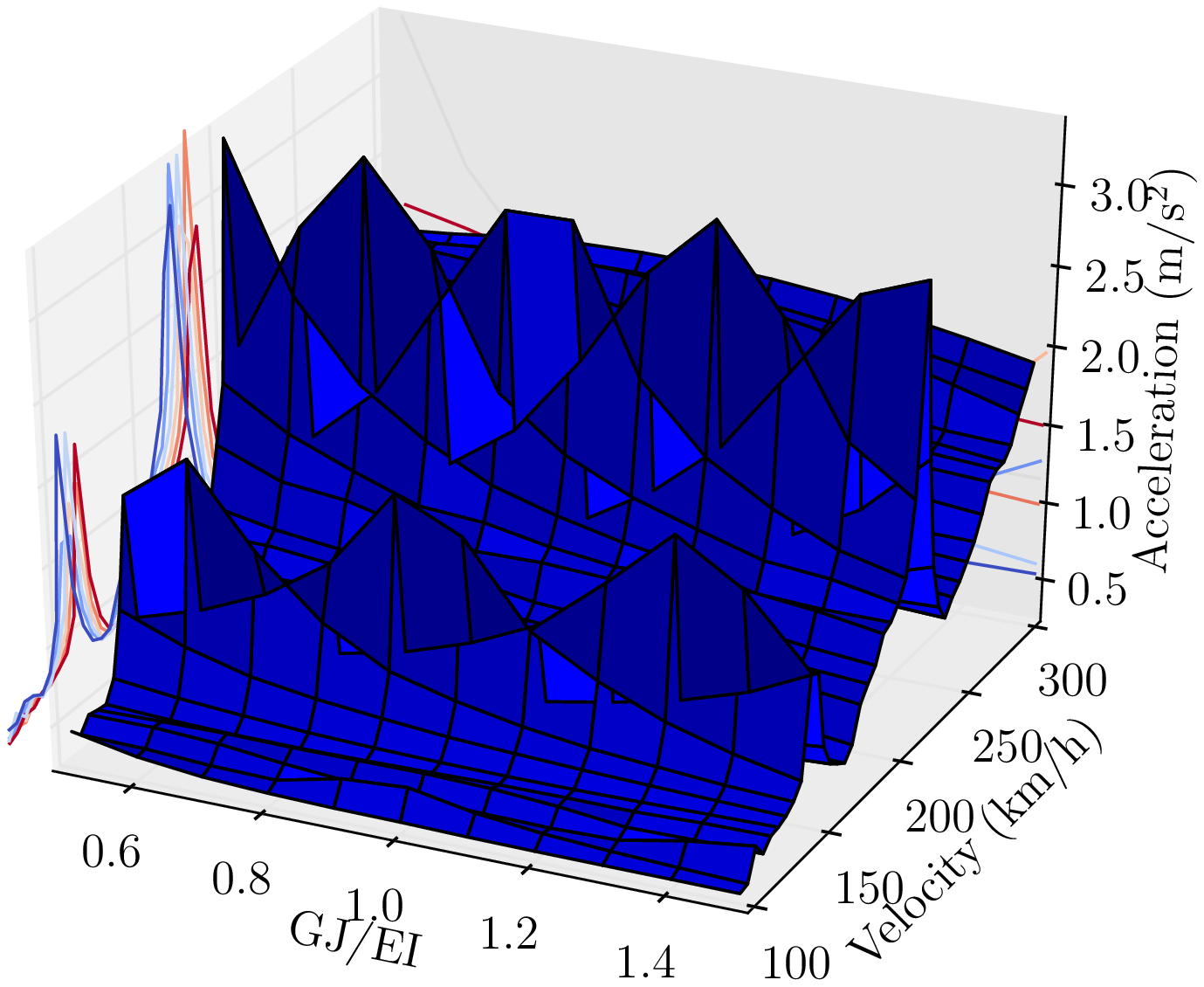}}
	\caption{Effect of torsional to flexural stiffness ratio on the maximum dynamic responses for a skew angle of 30$^{\circ}$: a) displacement, b) acceleration}
	\label{fig.influence_GJ_skew30}
\end{figure}

For this study, the torsional stiffness ($GJ$) is changed with respect to the flexural stiffness ($EI$) such that the ratio between $GJ$ and $EI$ varies in a range from 0.5 to 1.5. Figure \ref{fig.influence_GJ} shows the variation of the maximum dynamic responses at the mid-span as a function of the torsional to flexural stiffness ratio. It can be observed that the maximum vertical displacement increases slightly as the ratio increases, while the maximum acceleration is barely changed. It should be noted that the skew angle used for this study is constant and is 10$^{\circ}$. This skew angle is in a range from 0$^{\circ}$ to 15$^{\circ}$ in which the skewness has small influence on the dynamic response of the bridge as mentioned in the preceding section and shown in Fig. \ref{fig.influence_skewness1}. As a result of this, the torsional stiffness does not have a pronounced influence in the vertical deflection for small skew angles. For larger skew angle, 30$^{\circ}$ for example, the torsional stiffness has a noticeable effect on the maximum vertical displacement, as shown in Fig. \ref{fig.influence_GJ_skew30_1}. The maximum acceleration is almost completely unaffected by the torsional stiffness for both skew angles selected (see Fig. \ref{fig.influence_GJ_2} and \ref{fig.influence_GJ_skew30_2}). 

\subsection{Effect of the span length}

In this part of the paper, the influence of the span length on the dynamic response of the simply-support skew bridge is carried out. The span length is changed from 15 m to 35 m in increment of 5 m. In order to obtain a consistent comparison between the results obtained from the parametric study, the cross section of the bridge is redesigned for each span length, using the design criteria that the ratio between the depth of the cross section (h) and the span length (L) is constant and is 1/14. This ratio is usually applied in the railway bridge design. The depth of the cross section will be changed with the bridge's length. The other dimensions of the cross section are considered as unmodified.  The basic properties of the cross section needed for the parametric study are listed in Table \ref{tab.3}. 

\begin{figure}[h!]
	\centering
	\subfigure[]{\label{fig.frequencies}\includegraphics[width=0.48\textwidth]{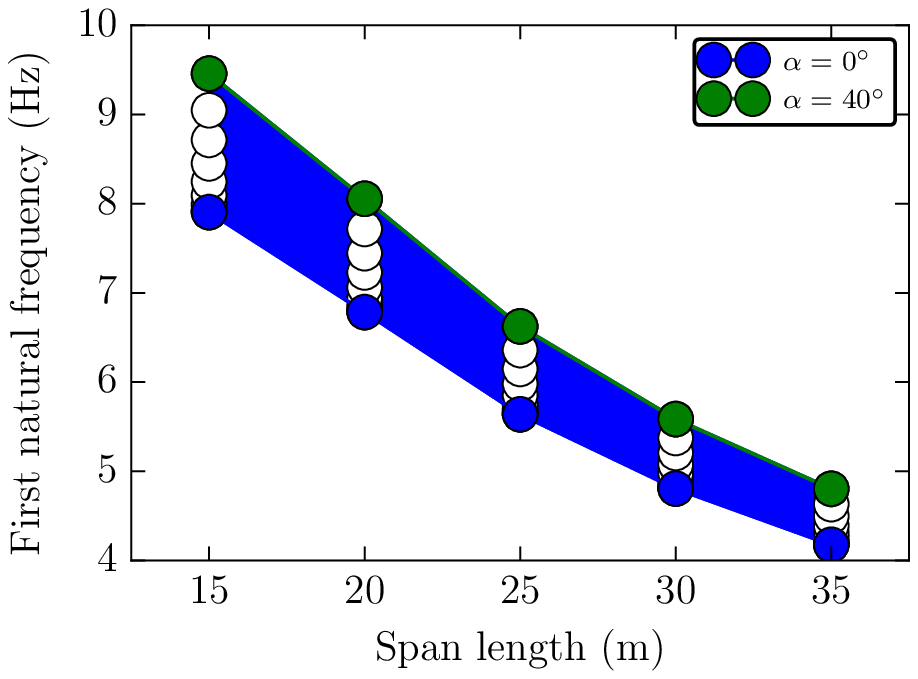}}
	\subfigure[]{\label{fig.variation}\includegraphics[width=0.48\textwidth]{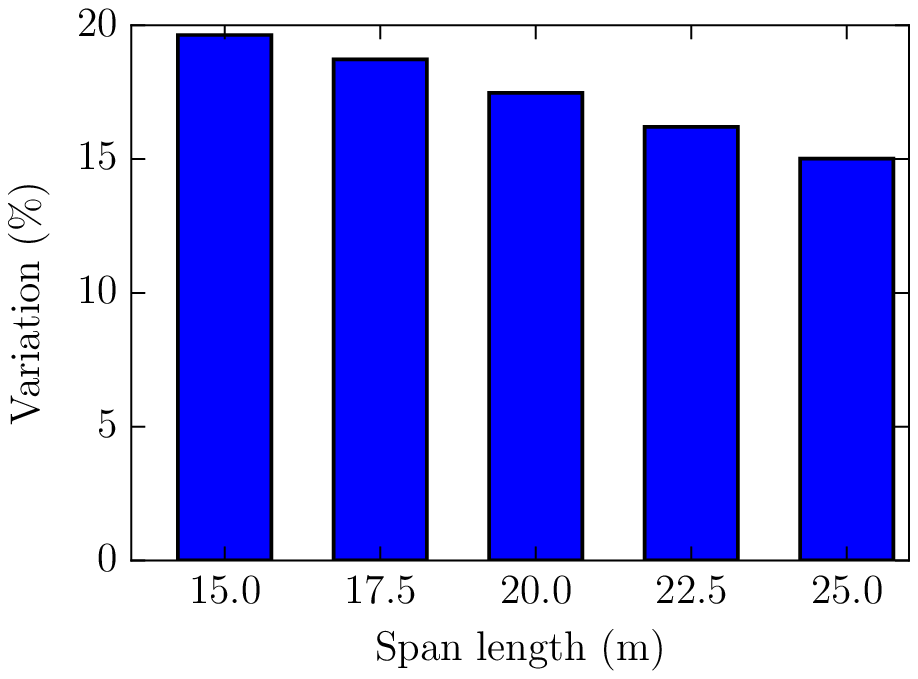}}
	\caption{Influence of the span length on the natural frequency of the simply-supported skew bridge: a) first natural frequency, b) variation of frequency}
	\label{fig.influence_L_frequency}
\end{figure}

\begin{table}[h!]
	\centering
	\caption{Principal properties of the bridge for the parametric study}
	\begin{tabular}{cccccc}
		L (m) & h (m) & $EI$ (N.m$^2$) & $GJ$ (N.m$^2$) & m (t/m) & h/L \\ \hline
		15.0 & 1.07 & 11.66e09 & 9.75e09 & 9.092 & 1/14 \\
		20.0 & 1.42 & 28.70e09 & 22.61e09 & 9.616 & 1/14 \\
		25.0 & 1.78 & 50.85e09 & 36.26e09 & 10.101 & 1/14 \\
		30.0 & 2.14 & 80.40e09 & 53.21e09 & 10.603 & 1/14 \\
		35.0 & 2.50 & 117.91e09 & 71.34e09 & 11.116 & 1/14 \\ \hline
	\end{tabular}
	\label{tab.3}
\end{table}
The first natural frequency corresponding to each span length is obtained and depicted in Fig. \ref{fig.frequencies} for different skew angles varying from 0$^{\circ}$ to 40$^{\circ}$ and the variation of magnitude of the first natural frequency between the skew angle of 0$^{\circ}$ and 40$^{\circ}$ for each span length is also obtained and shown in Fig. \ref{fig.variation}. It can be observed that the variation of frequency for each span length is generated by the skewness effect. This variation is greater when the span length is shorter and decreases almost linearly with span length. Therefore, it can be remarked that the span length decreases the skewness effect on the bridge in term of the natural frequency.  

\begin{figure}[h!]
	\centering
	\subfigure[]{\label{}\includegraphics[width=0.48\textwidth]{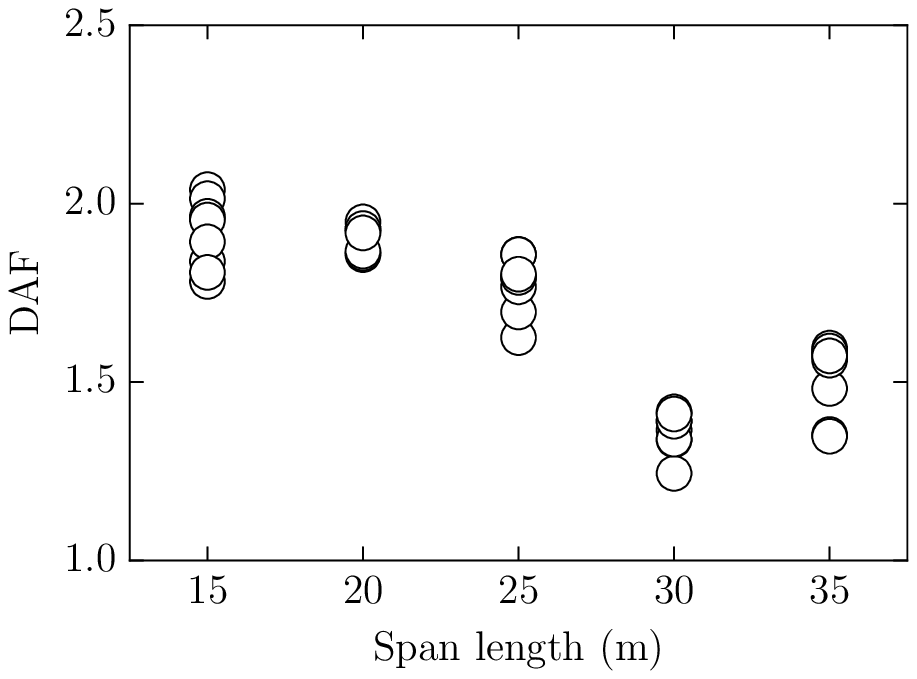}}
	\subfigure[]{\label{}\includegraphics[width=0.48\textwidth]{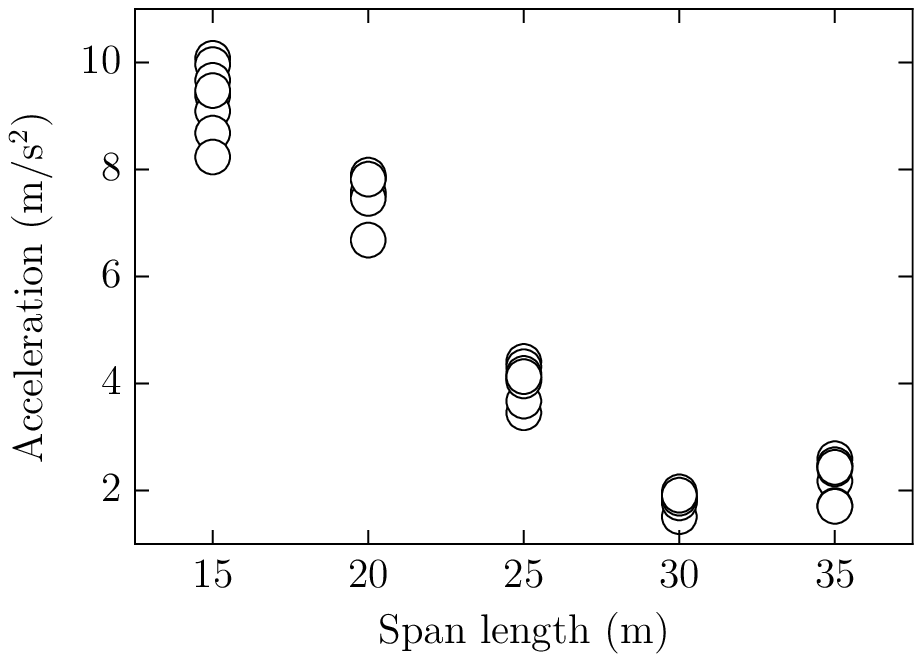}}
	\caption{Maximum dynamic responses at mid-span of the skew bridge at the peak corresponding to the second velocity of resonance for different skew angles: a) dynamic amplification factor of displacement, b) acceleration }
	\label{fig.influence_L}
\end{figure}

It is well known that the dynamic response of a bridge under the traffic loads depends on the properties of the vehicle traveling on the bridge and on the proper characteristics of the bridge. In this parametric study, the traffic loads are unmodified, but the characteristics of the bridge are changed with the span length. Therefore, the comparison of dynamic responses in term of displacement and acceleration in time-history at the determined train velocity is not consistent. For a consistent comparison, the peak corresponding to the second train velocity of resonance for each span length is compared, in particular, the dynamic amplification factor (DAF) of the vertical displacement and the maximum vertical acceleration at mid-span are used to compare and are depicted in Fig. \ref{fig.influence_L}. It can be observed that the DAF decreases as the span length increases. There is not a reduction of variation of magnitude of DAF of the displacement for different skew angles when the span length increases. However, the reduction of variation of magnitude of the maximum acceleration can be observed for different skew angles, for which it can be remarked that the span length reduces the skewness effect on the dynamic response of the bridge in term of the vertical acceleration.

\section{Conclusions}
In this paper, an analytical model for determining the dynamic response of the simply-supported skew bridge under the moving loads is presented and a simplified model is also proposed. The modal superposition technique is used in both models to decompose the differential equation of motions. The natural frequencies and mode shapes and the orthogonality relationship are determined from the boundary conditions. The time-dependent modal equations are solved by the exact integration, and therefore, the both models are highly accurate, robust and computationally efficient. The proposed models have been validated with results obtained from the FE models using the same modal superposition method. Furthermore, from the results obtained in this paper, the following conclusions are made:
\begin{itemize}
	\item The estimation of the train velocities of resonance proposed by \cite{CEN2003} is still valid for the simply-supported skew bridge.
	\item The grade of skewness of the bridge plays important role in the dynamic behavior of the bridge in term of the vertical displacement. The maximum vertical displacement decreases as the skew angle increases. The vibration of bridge in term of the vertical acceleration is hardly affected by the skewness.
	\item There is a critical skew angle from which the effect of the skewness is more noticeable. For the cross section used in the parametric study, the critical skew angle is 15$^{\circ}$.
	\item The torsional stiffness really has important influence on the vibration of the bridge in term of the vertical displacement when the skew angle is larger than the critical skew angle. The vertical acceleration is unaffected by the torsional stiffness.
	\item The span length reduces the skewness effect on the dynamic behavior of the skew bridge in term of the natural frequency and acceleration.
\end{itemize}
\appendix
\section{Parameters for the exact integration}
\begin{align}
	A &= e^{-\xi_n\omega_n\Delta t}\left( \frac{\xi_n}{\sqrt{1-\xi_n^2}}\sin\omega_D\Delta t+\cos\omega_D\Delta t\right) \\
	B &=  e^{-\xi_n\omega_n\Delta t} \left(\frac{1}{\omega_D}\sin\omega_D\Delta t \right) \\
	C &= \frac{1}{\omega_n^2}\left\{\frac{2\xi_n}{\omega_n \Delta t} + e^{-\xi_n\omega_n\Delta t}\left[ \left(\frac{1-2\xi_n^2}{\omega_D\Delta t}-\frac{\xi_n}{\sqrt{1-\xi_n^2}}\right) \sin\omega_D\Delta t - \left(1+\frac{2\xi_n}{\omega_n \Delta t} \right) \cos\omega_D\Delta t \right] \right\} \\
	D &= \frac{1}{\omega_n^2}\left[1-\frac{2\xi_n}{\omega_n \Delta t}+e^{-\xi_n\omega_n\Delta t} \left(\frac{2\xi_n^2-1}{\omega_D \Delta t} \sin\omega_D\Delta t + \frac{2\xi_n}{\omega_n \Delta t}\cos\omega_D \Delta t \right)\right] \\
	A' &= -e^{-\xi_n\omega_n\Delta t}\left( \frac{\omega_n}{\sqrt{1-\xi_n^2}}\sin\omega_D\Delta t\right) \\
	B' &=  e^{-\xi_n\omega_n\Delta t} \left(\cos\omega_D\Delta t-\frac{\xi_n}{1-\xi_n^2}\sin\omega_D\Delta t \right) \\
	C' &= \frac{1}{\omega_n^2}\left\{-\frac{1}{\Delta t} + e^{-\xi_n\omega_n\Delta t}\left[ \left(\frac{\omega_n}{\sqrt{1-\xi_n^2}}+\frac{\xi_n}{\sqrt{1-\xi_n^2}}\right) \sin\omega_D\Delta t +\frac{1}{\Delta t}\cos\omega_D\Delta t \right] \right\} \\
	D' &= \frac{1}{\omega_n^2 \Delta t}\left[1-e^{-\xi_n\omega_n\Delta t} \left(\frac{\xi_n}{\sqrt{1-\xi_n^2}}\sin\omega_D\Delta t + \cos\omega_D \Delta t \right)\right] 
\end{align}
where $\omega_D = \omega_n\sqrt{1-\xi_n^{2}}$
\section*{Acknowledgement}
The authors are grateful to the support of MINECO of Spanish Government through
the project EDINPF (Ref. BIA2015-71016-R) and to the support provided by the Technical University of Madrid, Spain. 
\bibliography{reference}

\begin{thebibliography}{10}
\expandafter\ifx\csname url\endcsname\relax
  \def\url#1{\texttt{#1}}\fi
\expandafter\ifx\csname urlprefix\endcsname\relax\def\urlprefix{URL }\fi
\expandafter\ifx\csname href\endcsname\relax
  \def\href#1#2{#2} \def\path#1{#1}\fi

\bibitem{Kollbrunner1969}
C.~F. Kollbrunner, K.~Basler, Torsion in Strucutres: an engineering approach,
  Springer-Verlag, Berlin, 1969.

\bibitem{Manterola2006}
J.~Manterola, Bridges: design, calculation and construction (in Spanish),
  Colegio de Ingenieros de Caminos, Canales y Puertos, Madrid, Spain, 2006.

\bibitem{Ghobarah1974}
A.~A. Ghobarah, W.~K. Tso, {Seismic analysis of skewed highway bridges with
  intermediate supports}, Earthquake Engineering {\&} Structural Dynamics 2~(3)
  (1974) 235--248.

\bibitem{Maragakis1987}
E.~A. Maragakis, P.~C. Jennings, {Analytical models for the rigid body
  motions}, Earthquake Engineering {\&} Structural Dynamics 15~(January) (1987)
  923--944.

\bibitem{Wakefield1991}
R.~R. Wakefield, A.~S. Nazmy, D.~P. Billington, {ANALYSIS OF SEISMIC FAILURE IN
  SKEW RC BRIDGE}, Journal of Structural Engineering 117~(3) (1991) 972--986.

\bibitem{Meng2000}
J.~Y. Meng, E.~M. Lui, {Seismic analysis and assessment of a skew highway
  bridge}, Engineering Structures 22~(11) (2000) 1433--1452.

\bibitem{Meng2001}
J.~Y. MENG, E.~M. LUI, Y.~LIU, {Dynamic Response of Skew Highway Bridges},
  Journal of Earthquake Engineering 5~(2) (2001) 205--223.

\bibitem{Nielson2007}
B.~G. Nielson, R.~DesRoches, {Analytical seismic fragility curves for typical
  bridges in the central and southeastern United States}, Earthquake Spectra
  23~(3) (2007) 615--633.

\bibitem{Abdel-Mohti2008}
A.~Abdel-Mohti, G.~Pekcan, {Seismic response of skewed RC box-girder bridges},
  Earthquake Engineering and Engineering Vibration 7~(4) (2008) 415--426.

\bibitem{Kaviani2012}
P.~Kaviani, F.~Zareian, E.~Taciroglu, {Seismic behavior of reinforced concrete
  bridges with skew-angled seat-type abutments}, Engineering Structures 45
  (2012) 137--150.

\bibitem{Yang2015}
C.~S.~W. Yang, S.~D. Werner, R.~DesRoches, {Seismic fragility analysis of
  skewed bridges in the central southeastern United States}, Engineering
  Structures 83 (2015) 116--128.

\bibitem{Meng2002}
J.-Y. Meng, E.~M. Lui, {Refined stick model for dynamic analysis of skew
  highway bridges}, Journal of Bridge Engineering 7~(3) (2002) 184--194.

\bibitem{Nouri2012}
G.~Nouri, Z.~Ahmadi, {Influence of Skew Angle on Continuous Composite Girder
  Bridge}, Journal of Bridge Engineering 17~(4) (2012) 617--623.

\bibitem{Deng2015}
Y.~Deng, B.~M. Phares, L.~Greimann, G.~L. Shryack, J.~J. Hoffman, {Behavior of
  curved and skewed bridges with integral abutments}, Journal of Constructional
  Steel Research 109 (2015) 115--136.

\bibitem{Mallick2015}
M.~Mallick, P.~Raychowdhury, {Seismic analysis of highway skew bridges with
  nonlinear soil-pile interaction}, Transportation Geotechnics 3 (2015) 36--47.

\bibitem{Bishara1993}
A.~G. Bishara, M.~C. Liu, N.~D. El-Ali, {Skew I-Beam Composite Bridges},
  Journal of Structural Engineering 119~(2) (1993) 399--419.

\bibitem{Helba1995}
A.~Helba, J.~B. Kennedy, {Skew composite bridges ultimate load}, Canadian
  Journal of Civil Engineering 22 (1995) 1092--1103.

\bibitem{Khaloo2003}
A.~R. Khaloo, H.~Mirzabozorg, {Load Distribution Factors in Simply Supported
  Skew Bridges}, Journal of Bridge Engineering 8~(4) (2003) 241--244.

\bibitem{Menassa2007}
C.~Menassa, M.~Mabsout, K.~Tarhini, G.~Frederick, {Influence of skew angle on
  reinforced concrete slab bridges}, Journal of Bridge Engineering 12~(2)
  (2007) 205--214.

\bibitem{Ashebo2007}
D.~B. Ashebo, T.~H.~T. Chan, L.~Yu, {Evaluation of dynamic loads on a skew box
  girder continuous bridge Part I: Field test and modal analysis}, Engineering
  Structures 29~(6) (2007) 1052--1063.

\bibitem{He2012}
X.~H. He, X.~W. Sheng, A.~Scanlon, D.~G. Linzell, X.~D. Yu, {Skewed concrete
  box girder bridge static and dynamic testing and analysis}, Engineering
  Structures 39 (2012) 38--49.

\bibitem{Chopra2012}
A.~K. Chopra, Dynamics of Structures: Theory and Applications to Earthquake
  Engineering, 4th Edition, Prentice Hall, 2012.

\bibitem{feap}
R.~Taylor, \href{http://www.ce.berkeley/feap}{{FEAP}-finite element analysis
  program} (2014).
\newline\urlprefix\url{http://www.ce.berkeley/feap}

\bibitem{CEN2003}
CEN, {EN 1991-2:2003 Actions on Structures - Part 2: Traffic loads on bridges},
  rue de Stassart, 36B-1050 Brussels (2003).

\end{thebibliography}
\end{document}